\documentclass{PoS}
\PoS{PoS(HEP2005)325}
\usepackage{epsfig}

\newcommand{\ee}{\mbox{${\mathrm{e}}^+ {\mathrm{e}}^-$}}

\newcommand{\tautau}{\ensuremath{\tau^+\tau^-}}

\newcommand{\bb}         {\ensuremath{\mathrm{b}\bar{\mathrm{b}}}}

\newcommand{\mHone}         {\mbox{$m_{\mathrm{H}_{1}}$}}
\newcommand{\mHtwo}         {\mbox{$m_{\mathrm{H}_{2}}$}}

\newcommand{\Hone}{\ensuremath{\mathrm{H}_1}}
\newcommand{\Htwo}{\ensuremath{\mathrm{H}_2}}

\newcommand {\Ao}        {\ensuremath{\mathrm{A}}}
\newcommand {\ho}        {\ensuremath{\mathrm{h}}}
\newcommand {\Zo}        {\ensuremath{\mathrm{Z}}}

\newcommand{\mHpm}{\mbox{$m_{\mathrm{H}^{\pm}}$}}

\newcommand{\tanb}       {\mbox{$\tan\beta$}}

\newcommand{\sqrts}     {\mbox{$\sqrt{s}$}}

\newcommand{\ra}        {\mbox{$\rightarrow$}}   % \to can be used as well

\title{Search for CP Violating neutral Higgs bosons in the MSSM at LEP}
\ShortTitle{Search for CP-violating neutral Higgs bosons in the MSSM at LEP}
\author{\speaker{Philip Bechtle}\thanks{For the LEP collaborations}\\        
  Stanford Linear Accelerator Center, 2575 Sand Hill Road, Menlo Park, CA 94025, USA\\        
  E-mail: \email{bechtle@slac.stanford.edu}}

\abstract{
  
  The LEP collaborations ALEPH, DELPHI, L3 and OPAL have searched for
  the neutral Higgs bosons which are predicted within the framework of
  the Minimal Supersymmetric Standard Model (MSSM). The data of the
  four collaborations are statistically combined and show no signicant
  excess of events which would indicate the production of Higgs
  bosons. The search results are thus used to set upper bounds on the
  cross sections of various Higgs-like event topologies and limits on
  MSSM benchmark models, including CP-conserving and CP-violating
  scenarios.  Here, the limits on the model parameters of the
  CP-violating benchmark scenario CPX and derivates of this scenario
  are shown. 

}

\FullConference{International Europhysics Conference on High Energy Physics\\
                 July 21st - 27th 2005\\
                 Lisboa, Portugal}

\begin{document}

%
% ---------------------------------------
%
\section{Introduction}

It is generally assumed that the Higgs mechanism~\cite{Higgs:1964pj}
is responsible for the breaking of electroweak symmetry and for the
generation of elementary particle masses, both in the Standard Model
and in one of its most popular extensions, Supersymmetry
(SUSY)~\cite{Wess:1974tw}. In the implementation of SUSY with
minimal additional particle content, the Minimal Supersymmetric
Standard Model (MSSM), two Higgs doublets are needed. Hence, three
neutral Higgs bosons and a pair of charged Higgs bosons can be
searched for.

Most of the experimental investigations carried out in the past at LEP
and elsewhere were interpreted in MSSM scenarios which assume CP
conservation in the Higgs sector. However, CP violation in the Higgs
sector cannot be a priori excluded~\cite{Pilaftsis:1999qt}. Scenarios
with CP violation are theoretically appealing since they provide one
of the ingredients needed to explain the observed cosmic
matter-antimatter asymmetry.  In the MSSM, however, substantial
CP-violating effects can be induced by radiative corrections,
especially from third generation scalar-quarks.  In the CP-violating
scenario the three neutral Higgs mass eigenstates, H1, H2 and H3, are
mixtures of CP-even and CP-odd fields. The Higgs boson production and
decay properties may therefore be widely different and, consequently,
the experimental exclusions published so far for the CP-conserving
MSSM scenario may be invalidated by CP-violating effects.

In this paper, the results from the combination of the Higgs boson
searches of the four LEP collaborations~\cite{AllSearches} at
$\sqrts=91-209\,\mathrm{GeV}$ in model-independent cross-section
limits on various MSSM-Higgs-like topologies and in exclusion of
CP-violating MSSM benchmark scenarios are presented.

\section{Higgs Boson Searches at LEP}

In the CP-conserving MSSM, the two dominant production mechanisms
Higgsstrahlung ($\ee\ra\ho\Zo$,
$\sigma_{\ho\Zo}\propto\sin^2(\beta-\alpha)$) and pair production
($\ee\ra\ho\Ao$, $\sigma_{\ho\Ao}\propto\cos^2(\beta-\alpha)$) are
complementary and ensure the coverage of the whole kinematically
accessible plane of Higgs boson masses, because for large
$\cos^2(\beta-\alpha)$ the two Higgs bosons $\ho$ (CP-even) and $\Ao$
are close to each other in mass.

\begin{figure}[t]
  \begin{center}
    \epsfig{file=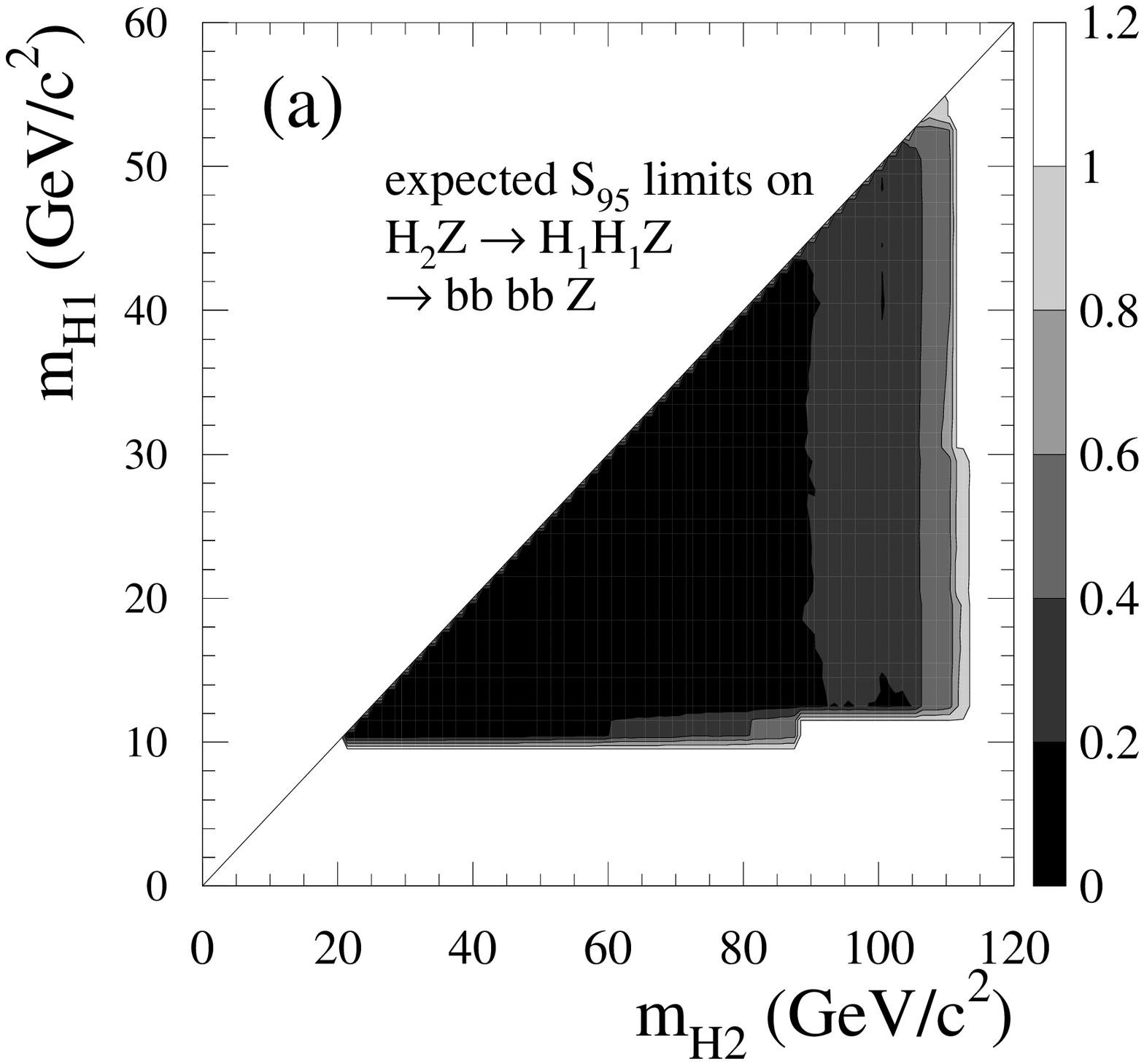, width=0.4\textwidth}
    \epsfig{file=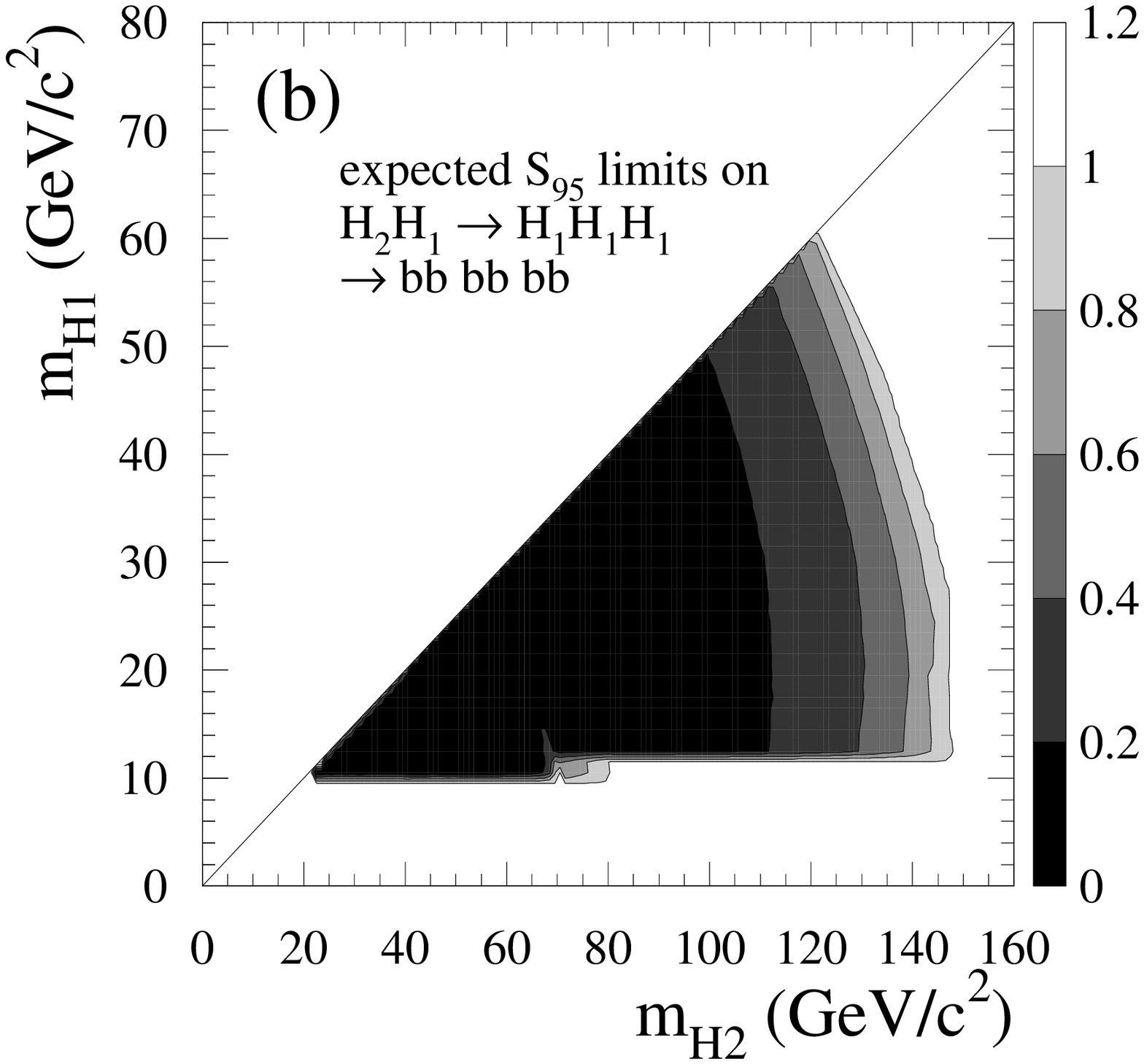, width=0.4\textwidth}\vspace{-7mm}
  \end{center}
  \caption{{\sl Model independent limits on $\sigma\times\mathrm{BR}$ relative 
      to the nominal 2HDM cross-sections for $\sin^2(\beta-\alpha)=1$
      in (a) and $\cos^2(\beta-\alpha)=1$ in (b).}}\label{fig:ModIndepLimits}
\end{figure}

In the CP-violating MSSM, the experimental coverage of the mass plane
is lost, since first all three Higgs bosons can be produced in
Higgsstrahlung (and hence the direct complementary of two modes is
lost), and second, because there can be large mass differences between
$\mHone$ and $\mHtwo$ over the whole parameter space. Additionally,
the cascade decay $\Htwo\ra\Hone\Hone$ is dominant in large areas of
the parameter space. Hence, the coverage of non-diagonal pair
production mechanisms and the coverage of cascade decays is crucial
for the experimental access to the CP-violating models. This is shown
in Fig.~\ref{fig:ModIndepLimits}. In (a), the 95\,\% confidence level
(CL) exclusion limits on $\sigma\times\mathrm{BR}$ in the process
$\ee\ra\Htwo\Zo\ra\Hone\Hone\Zo\ra\bb\bb\Zo$ relative to the nominal
SM cross-section is shown. For $\mHtwo$ up to 105\,GeV and all
$\mHone$, models which predict a $\sigma\times\mathrm{BR}$ value of
more than 40\,\% of the SM cross-section can be excluded. In (b), the
coverage for the process
$\ee\ra\Htwo\Hone\ra\Hone\Hone\Hone\ra\bb\bb\bb$ is shown, with limits
relative to the nominal pair-production cross-section with
$\cos^2(\beta-\alpha)=1$.

\section{MSSM Models with Additional CP Violation}\label{sec:theory}

The benchmark model used in the combination of the LEP data is the CPX
scenario~\cite{Carena:2000ks}. It is characterized by large mixing
between CP-even and CP-odd states in the mass eigenstates. The
CP-even/CP-odd mixing ${\cal M}_{SP}^2$ is characterized by 
\begin{equation}
{\cal M}_{SP}^2\propto
\frac{m_{t}^4}{v^2}\frac{\Im({A_{t,b}\mu})}{M_{SUSY}^2}.\label{eqn:mixing}
\end{equation}
Therefore,
large values of the top quark mass $m_{t}$, the Higgsino mixing
parameter $\mu$ and the imaginary part of the trilinear couplings in
the stop and sbottom sector {$A_{t,b}$}, coupled with a not too large
scale of the squark masses $M_{SUSY}$ is chosen. Effects of the
variation of these parameters are studied. Detailed calculations on
the two-loop order~\cite{Heinemeyer:1998yj} or on the one-loop
renormalization-improved order~\cite{Carena:2000ks} are used to
calculate the model predictions. 

\begin{figure}[t]
  \begin{center}
    \epsfig{file=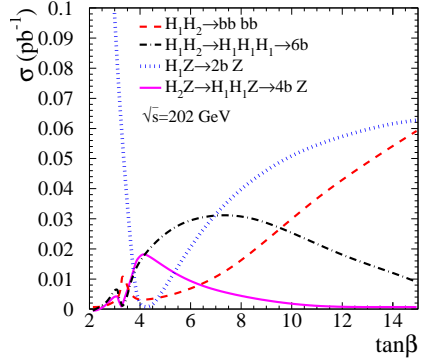,
      width=0.4\textwidth}\vspace{-7mm}
  \end{center}
  \caption{{\sl Model predictions for $\sigma\times\mathrm{BR}$ values of dominant Higgs boson 
    production mechanisms in the CPX scenario for
    $30\,\mathrm{GeV}<\mHone<40\,\mathrm{GeV}$.}}\label{fig:CPXmodel}
\end{figure}

The resulting predictions for selected processes in the CPX scenario
are shown in Fig.~\ref{fig:CPXmodel} for lightest Higgs masses of
$30\,\mathrm{GeV}<\mHone<40\,\mathrm{GeV}$. For low $\tanb=v_2/v_1$,
the SM-like production mechanism $\Hone\Zo\ra(\bb,\tautau)\Zo$ is
dominant and has a large production cross-section. For intermediate
$\tanb$, however, all production cross-sections are reduced, since the
kinematically accessible $\Hone$ decouples from the $\Zo$, hence no
Higgsstrahlung occurs, and since $\mHtwo\approx110\,\mathrm{GeV}$ is
close to the kinematic limit. Additionally, the experimentally more
difficult cascade decay $\Htwo\ra\Hone\Hone$ becomes dominant. For
large $\tanb$ the production cross-sections increase and finally
$\Hone\Htwo\ra\bb\bb$ becomes the dominant mode.

\section{Interpretation of the LEP Data in the CPV MSSM}

\begin{figure}[t]
  \begin{center}
    \epsfig{file=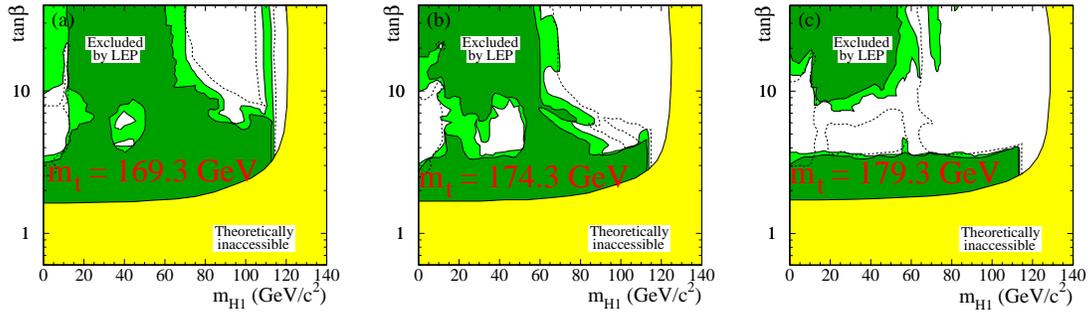,
      width=\textwidth}\vspace{-7mm}
  \end{center}
  \caption{{\sl Exclusion areas in the $(\tanb,\mHone)$ plane in the CPX 
      scenario for
      $m_{\mathrm{t}}=169.3,\,174.3,\,179.3\,\mathrm{GeV}$.
      Theoretically inaccessible regions are shown in yellow,
      experimentally excluded areas in light green
      ($\mathrm{CL}=95\,\%$) and dark green ($\mathrm{CL}=99.7\,\%$).}}\label{fig:CPXmtop}
\end{figure}

\begin{figure}[t]
  \begin{center}
    \epsfig{file=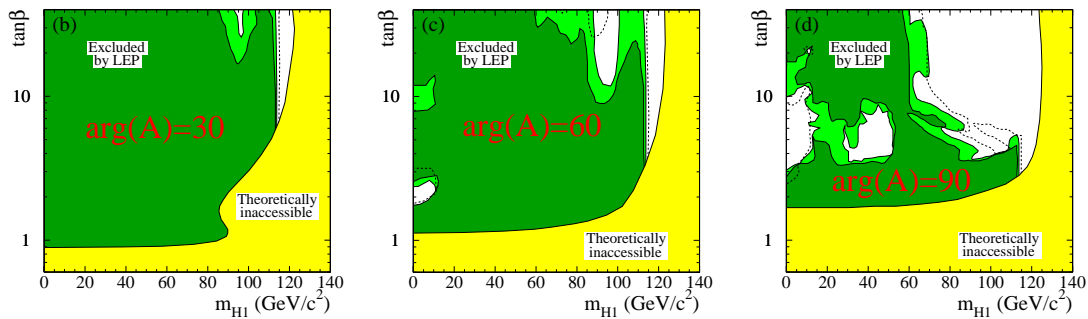,
      width=\textwidth}\vspace{-7mm}
  \end{center}
  \caption{{\sl Exclusion areas in the $(\tanb,\mHone)$ plane for different values 
      of the phase $\arg{A}$ of the trilinear coupling parameters in
      the stop and sbottom sector.}}\label{fig:CPXargA}
\end{figure}

The statistical combination of all Higgs boson searches from all four
LEP collaborations uses the modified frequentist approach as
implemented in~\cite{Junk:1999kv}. The result of this combination
shows no statistically significant excesses of the data over the
expected background. Hence, limits on the parameter space are
computed~\cite{LEPcombination}. These limits are shown in
Fig.~\ref{fig:CPXmtop} and \ref{fig:CPXargA} for the CPX scenario. In
each case, the full set of MSSM parameters is fixed to the values
chosen for the scenario (as given in
\cite{Carena:2000ks,LEPcombination}), apart from $\tanb$ and the
charged Higgs boson mass $\mHpm$, which are scanned. The result is
then shown in the $\tanb,\mHone$ projection.

In Fig.~\ref{fig:CPXmtop} the results in the CPX scenario are shown
for different top quark masses $m_{t}$. The present experimental value
of $m_{t}=172.7\,\mathrm{GeV}$~\cite{mtop} lies between the values
used for Fig.~\ref{fig:CPXmtop}~(a) and (b). In all cases, the
reduction of production cross-sections for intermediate $\tanb$
described in Section~\ref{sec:theory} causes unexcluded regions for
low values of the lightest Higgs boson mass $\mHone$. No absolute
limit on $\mHone$ can be set. For larger values of $m_{t}$ this effect
increases, since $m_{t}$ strongly influences the mixing of the mass
eigenstates (see (\ref{eqn:mixing})) and increases the mass
splitting between $\mHone$ and $\mHtwo$, hence further decreasing the
production cross-sections for intermediate $\tanb$.

The effect of unexcluded regions in the parameter space for low
$\mHone$ is clearly connected to the CP-violating imaginary phase of
the trilinear couplings $A_{t,b}$. This is shown in
Fig.~\ref{fig:CPXargA}. Only for large phases (and hence large mixings
in (\ref{eqn:mixing})) the effect of large inaccessible regions is strong.

\section{Conclusions}

The results from neutral Higgs bosons searches in the context of the MSSM
described in this paper are based on data collected by the four LEP
collaborations, ALEPH, DELPHI, L3 and OPAL at
$\sqrts=91-209\,\mathrm{GeV}$. No significant excess of data over the expected
backgrounds has been found. From these results, upper bounds are derived for
the cross sections of a number of Higgs event topologies. These upper
bounds cover a wide range of Higgs boson masses and are typically much lower
than the largest cross sections predicted within the MSSM framework.
In the CP-violating benchmark scenario CPX and the variants which have
been studied, the combined LEP data show large unexcluded domains,
down to the smallest masses; hence, no absolute limits can be set for
the Higgs boson masses. On the other hand, $\tanb$ can be restricted
to values larger than 2.9 for $m_{t}=174.3\,\mathrm{GeV}$. While the
excluded mass domains vary considerably with $m_{t}$, the bound in $\tanb$
is barely sensitive to the precise choice of the top quark mass.

%
% --------------------------------------------------------------------
%

\end{document}